# Shallow shear-wave (SV) reflection surveying with distributed acoustic sensing in Zuidbroek, the Netherlands

Dongyufu Zhang[1], Guy Drijkoningen[1]

(1. Department of Geoscience & Engineering, Delft University of Technology, The Netherlands)

## Introduction

High-density and high-quality seismic data acquisition have been one of the key factors driving the continuous advancement of seismology. Using optical fibres is a new technology that is emerging in the field of exploration (and earthquake) seismology, and is known as Distributed Acoustic Sensing (DAS). A DAS system is capable of detecting acoustic waves by quantifying the Rayleigh backscattering of a laser pulse. This process measures the changes in the fibre's axial strain that result from elastic vibrations (Rossi *et al.*, 2022). It not only has the advantages of ordinary optical fibre sensing technology (anti-electromagnetic interference, insulation, corrosion resistance, etc.), but can also realize long-distance real-time quantitative monitoring of dynamic strains along the optical fibre.

In recent years, DAS technology has been applied in well bores for vertical seismic profiling (Mateeva *et al.*, 2014), time-lapse monitoring (Mateeva *et al.*, 2017), shallow-structure imaging (Dou *et al.*, 2017), etc. For surface acquisition, Spikes *et al.*, (2019) acquired near-surface reflection data by deploying the fibres on the surface, but few studies on shallow shear-wave reflection surveys exist, let alone for DAS data. Shear waves have the characteristic that their propagation is mainly related to the rock or soil framework and thus hardly sensitive to the filling of pores (air or water), whereas the propagation of compressional waves is greatly affected by these fillings. This is particularly an issue in shallow investigations, as a result, imaging with shear waves is (much) less hampered by the heterogeneity caused by the different pore fillings than with compressional waves. Also shear waves give results that can be more directly related to the applications in geotechnical engineering, namely the strengths of soils.

In the province of Groningen in the Netherlands, a shear-wave (SV) reflection survey was carried out using a DAS system and horizontal-geophone system respectively, in which the fibre-optic cables included straight and helically wound fibres. The aim of the survey is to compare the results obtained from the DAS and the geophone systems for shallow shear-wave seismic reflection surveying, specifically with regard to the DAS having a denser and different kind (continuously over a gauge length) of sampling.

## Background of Distributed Acoustic Sensing

Distributed Acoustic Sensing uses an optical fibre as a sensor and a laser device as an interrogator. The interrogator excites a series of laser pulses into the optical fibre and a small part of the incident light will be reflected back through scattering to the sensing unit. By recording the phase of back-scattering and analysing the difference, the axial strain of the optical fibre can be determined at each point, so as to realize continuously distributed detection of acoustic signals. The interference pattern of the back-scattered light is subject to variation upon the deformation of the fibre, with the phases of the back-scattered light waves serving as the primary determinant (Kuvshinov, 2016). Notably, the phases are linked with the optical path length. The optical path change caused by fibre deformation is given by:

$$\frac{\Delta L_0}{L_0} = e_L - \frac{\Delta v_{ph}}{v_{ph}} = \frac{\Delta d}{d} - \frac{\Delta v_{ph}}{v_{ph}} \qquad (1)$$

where $L_0$ is the optical path length, $v_{ph}$ is the phase velocity of light waves in the fibre, $d$ is the light travel distance in the zone of interference effects, $e_L$ is the fibre strain, and $\Delta L_0$, $\Delta v_{ph}$ and $\Delta d$ are the changes of the earlier-mentioned parameters.

**Field Experiment Setup**

The field experiment was conducted near the town of Zuidbroek in the province of Groningen in the Netherlands. A seismic shear-wave (SV) reflection survey was performed using both a DAS system and a geophone-based (Geode©) system, which were used to acquire 2D seismic data.

The total length of the DAS fibre-optic cable (Fig 1b) used in this experiment was 4300 m, with the first 1000 m located indoors, the next 1000 to 2000 m on the surface, and the remaining 2000 to 4300 m in the target area. The DAS fibre cables were buried in a trench at a depth of 2 m and covered with the original ground composed mainly of sand and clay. The total optical fibre consists of both straight and helically wound fibres, fusion spliced together as one long fibre. We only focussed on the results of the straight fibre since helically wound fibres suppress shear-wave motion (Baird, 2020) so these latter were not useful for this type of survey. A FEBUS A1-R system was used as interrogator. The gauge length of the system was set at 2 m, with a pulse repetition frequency of 10 kHz and a channel spacing (spatial sampling resolution) of 80 cm. The interrogator was located in a farmer's house and connected to the buried cables through the surface cable from 1000 to 2000 m.

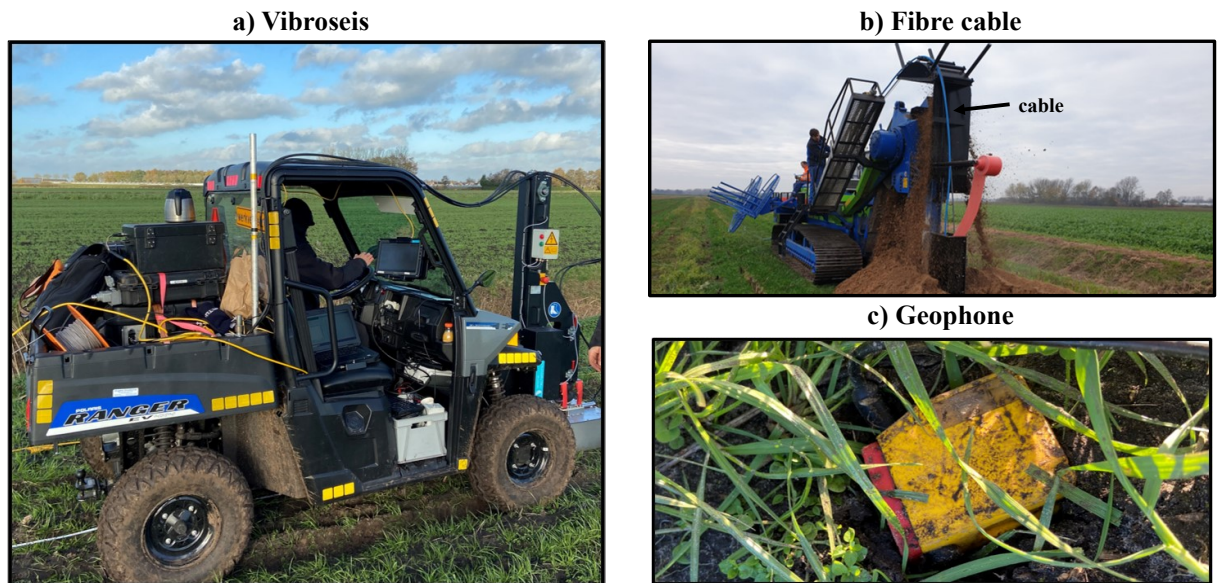

*Figure 1 (a) Electrically driven seismic vibrator (b) Installation of the fibre-optic cable (c) A horizontal geophone*

In the same location where the DAS cable was buried, a surface line consisting of 72 horizontal geophones (Fig 1c) was placed with a spacing of 2 m, giving a total length of 142 m. The Geode system was used to record the geophone data, and care was taken to ensure that the geophones were properly oriented and coupled to the soil. For the seismic source, an electrically driven vibratory source was utilized to generate a force of 900 N (Fig 1a), with a sweep length of 12 s and a frequency band of 3-100 Hz. Seismic data were acquired along a shooting line of 162 m, with intervals of 2 m between each shot position. Two shots were taken at each position for subsequent stacking.

**Seismic Data Processing**

Seismic data processing was conducted using Seismic Unix (SU), with the data processing sequence outlined in Table 1. For the DAS data, we first converted the FEBUS hdf5 file to the SU format and picked every trigger zero time. Given the sweep length of 12 s, the total listening time was 14 s so we selected a 14-second segment for each shot. We then correlated the data with the pilot, as shown in Fig 2, where the fibre-optic cables were divided into four parts (SF2, HWF2, SF1, HWF1), where we focussed on the SF2 part only. Then the coordinates were established and set, and a series of seismic data processes, including CMP-sort, filter, NMO, diversity stack, and time-migration, were performed. For the geophone data, the processing did not need any zero-time picking and selecting 14-s records since the Geode system is used as a triggered system. One important step for a proper comparison is that the geophone data needed to be converted to strain-rate data, and that is included in the processing sequence of the geophone data. The subsequent data processing flow was further the same as for the DAS data.

The results after processing are shown in Fig 3. Some shallow reflectors can be observed, and some migration smiles can be discerned, probably due to different coupling conditions of the optical fibres and the geophones. The resulting sections look quite similar on their structure.

However, differences can also be observed. The DAS data show better lateral continuity of the reflections than the geophone data, attributable to the more continuous sampling and the finer sampling resolution of 80 cm of the DAS data. Conversely, the spectra as shown in Fig.3 c and d reveal that the DAS data are characterized by a lower frequency content, with some maximum around 20 Hz, while the geophone data show a broader spectrum with no particular maximum.

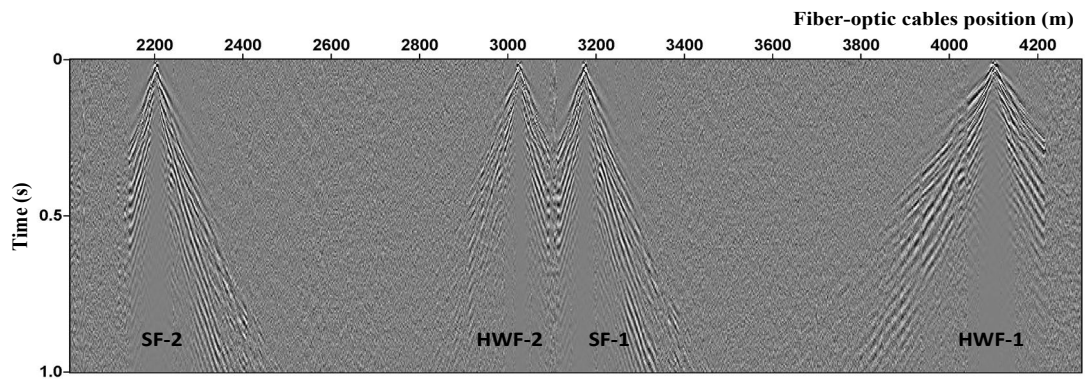

*Figure 2* Correlated single-shot data in time-distance domain. SF denoting Straight-Fibre data, and HWF denoting Helically Wound Fibre data. (Only SF2 data used in later processing.)

**Conclusions**

A field experiment was conducted in Zuidbroek, the Netherlands to compare the performance of a DAS and horizontal-geophone system for shear-wave (SV) reflection surveying. The data were subjected to processing for reflection imaging, including conversion of the geophone data to strain-rate data, to enable such a comparison on migrated-section level. Our findings indicate that DAS straight-fibre data shows a lower-frequency information content, but achieves better reflector continuity than the geophone data due to the more continuous and denser sampling with the DAS system.

**Acknowledgements**

This research has received funding from the European Research Council (ERC) under the European Union's Horizon 2020 research and innovation program (grant no. 742703).

*Table 1 Processing sequences for DAS and geophone data*

| Number | DAS data processing sequence | Geophone data processing sequence |
| --- | --- | --- |
| 1 | Convert hdf5 files to SU format | Convert SEG2 files to SU format |
| 2 | Pick zero time | |
| 3 | Select 14 seconds segments | |
| 4 | Correlate data with pilot | Correlate data with pilot |
| 5 | Select SF2 section from fibre line | |
| 6 | Vertical stack of 2 shots per shot point | Vertical stack of 2 shots per shot point |
| 7 | | Convert to strain-rate data |
| 8 | Set coordinates | Set coordinates |
| 9 | F-k filter to suppress surface waves | F-k filter to suppress surface waves |
| 10 | CMP-sort, NMO and diversity stack | CMP-sort, NMO and diversity stack |
| 11 | Time-Migration | Time-migration |

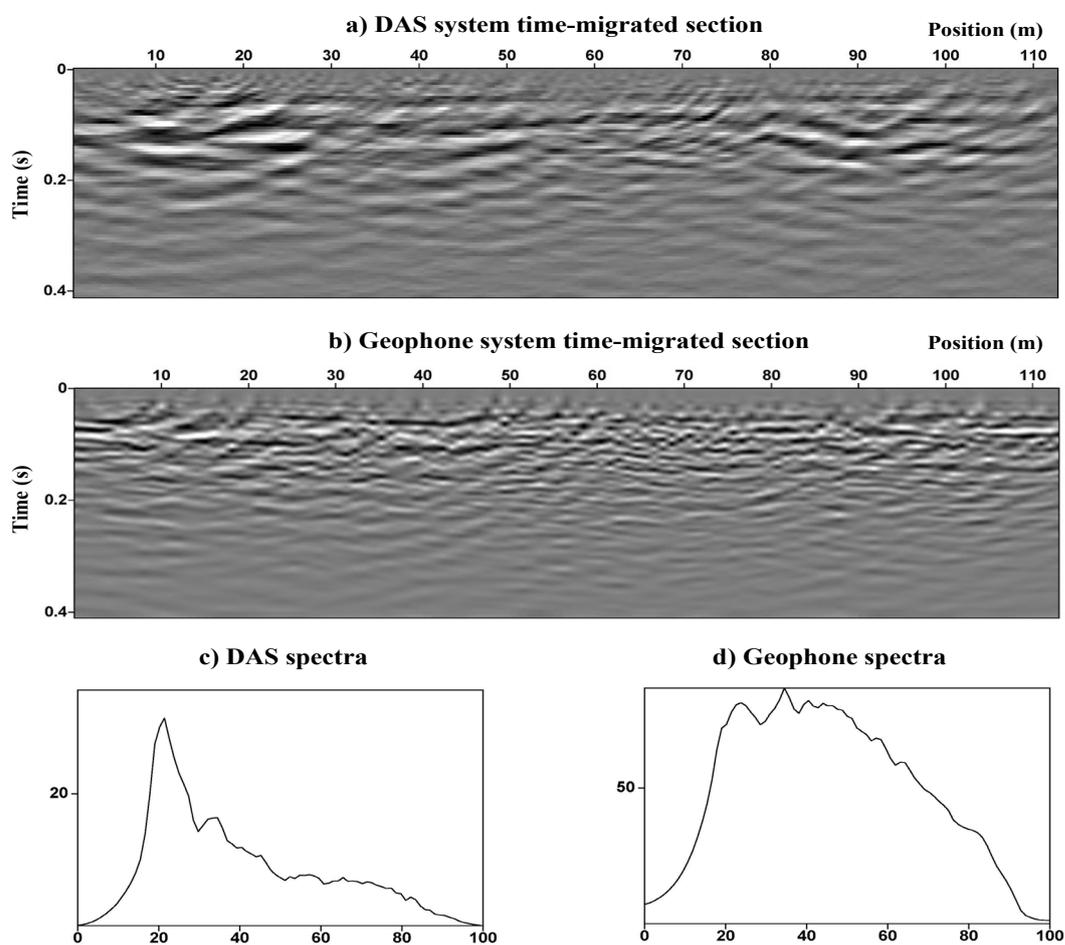

*Figure 3 (a) DAS system time-migrated section (b) Geophone system time-migrated section and their spectra for the DAS (c) and Geophone system (d).*